# A Decade of Scholarly Research on Open Knowledge Graphs

**Houcemeddine Turki**[*,κ,φ], **Abraham Toluwase Owodunni**[†,κ],
**Mohamed Ali Hadj Taieb**[*,κ], **René Fabrice Bile**[‡,κ] **and Mohamed Ben Aouicha**[*,κ]

[*]Data Engineering and Semantics Research Unit, University of Sfax, Sfax, Tunisia
turkiabdelwaheb@hotmail.fr, {mohamedali.hajtaieb, mohamed.benaouicha}@fss.usf.tn
[†]Masakhane, Abuja, Nigeria
owodunniabraham@gmail.com
[‡]National Polytechnic School of Maroua, University of Maroua, Maroua, Cameroon
bilerene@gmail.com
[κ]SisonkeBiotik Research Community, Johannesburg, South Africa
[φ]University of the People, Pasadena, California, United States of America

**Abstract**

The proliferation of open knowledge graphs has led to a surge in scholarly research on the topic over the past decade. This paper presents a bibliometric analysis of the scholarly literature on open knowledge graphs published between 2013 and 2023. The study aims to identify the trends, patterns, and impact of research in this field, as well as the key topics and research questions that have emerged. The work uses bibliometric techniques to analyze a sample of 4445 scholarly articles retrieved from Scopus. The findings reveal an ever-increasing number of publications on open knowledge graphs published every year, particularly in developed countries (+50 per year). These outputs are published in highly referred scholarly journals and conferences. The study identifies three main research themes: (1) knowledge graph construction and enrichment, (2) evaluation and reuse, and (3) fusion of knowledge graphs into NLP systems. Within these themes, the study identifies specific tasks that have received considerable attention, including entity linking, knowledge graph embedding, and graph neural networks.

## 1. Introduction

The past decade has witnessed remarkable growth in the use of open knowledge graphs, which has led to an increase in associated scholarly research (Hogan et al., 2021). Open knowledge graphs provide a structured representation of knowledge, making it easier to access and analyze information, and facilitating the development of intelligent applications (Hogan et al., 2021). During the last decade, several broad-coverage lexico-semantic knowledge resources (e.g., *Wikipedia* and *WordNet*) have been processed to create large-scale open knowledge graphs (Hogan et al., 2021). As a result, there has been a surge in interest in understanding the construction, evaluation, and integration of these fully structured databases. In this context, several literature surveys have been done to understand how open knowledge graphs are constructed (Hossain et al., 2020), represented (Cambria et al., 2021), enriched (Färber et al., 2017; Hogan et al., 2021), integrated (Mountantonakis and Tzitzikas, 2019), and validated (Paulheim, 2016). Several bibliometric studies have been conducted as well to study how scholarly research about knowledge graphs has evolved over the years (Gandon, 2018; Chen et al., 2021). However, these studies have not emphasized the evolution of open knowledge graphs and how they are currently used and developed. Open knowledge graphs are easily findable, accessible, interoperable, and reusable semantic resources that can provide lexical information in a variety of natural languages for free (Färber et al., 2017), making them extremely valuable resources in computational linguistics.

In this paper, we present a bibliometric study about scholarly research on open knowledge graphs between 2013 and 2022 based on *Scopus*, a large-scale bibliographic database maintained by Elsevier (Burnham, 2006). Our decision to utilize Scopus as the primary data source was driven by its status as the largest proprietary bibliographic database, renowned for its human curation and emphasis on verified, peer-reviewed research (Baas et al., 2020). The data quality in Scopus, known for its cleanliness and reliability, is superior to that of automatically generated databases like Google Scholar (Baas et al., 2020). While we acknowledge the potential value of including additional databases and qualitative analyses, our study was intentionally designed as an exploratory inquiry. Incorporating extensive qualitative methods such as expert interviews would have significantly broadened the scope and extended the duration of our research. Our primary objective was to provide a foundational, quantitative overview of the field, setting the stage for future research that could integrate these additional dimensions.

We begin by providing an overview of open knowl-edge graphs (Section 2). We restricted our analysis to Scopus because it is the largest human-curated bibliographic database only including quality semi-structured bibliographic information for peer-reviewed scholarly publications (Baas et al., 2020). Then, we specify our approach for analyzing the research outputs related to open knowledge graphs (Section 3). After that, we present how the scholarly production about the topic has evolved from the perspective of quantity, scholarly impact, country distribution, and source titles (Section 4). Later, we identify the main topics of the considered publications based on the analysis of their keywords and we study how these topics interacted together over ten years through the evaluation of their co-occurrence networks (Section 5). Subsequently, we discuss our results through comparison with the research findings of previous publications on this topic (Section 6). Finally, we draw conclusions for this output and provide future directions for open knowledge graph research (Section 7).

## 2. Open Knowledge Graphs

Knowledge graphs and ontologies are both semantic databases that represent information as statements in the form of triples: *Subject*, *Predicate*, and *Object*. Statements can either be relational (where both the subject and object are concepts) or non-relational (where the object is a string, a monolingual text, a value, a URL, a DateTime, or an external identifier) (Turki et al., 2019). The difference between knowledge graphs and ontologies is that the latter put semantic information about all its concepts in a single file, mostly in the Web Ontology Language (OWL) Format, while knowledge graphs store semantic knowledge about every concept in a separate file, mainly in the Resource Description Framework (RDF) Format (Fensel et al., 2020). This allows knowledge graphs to be more voluminous than ontologies and to include a large-scale variety of data, by contrast to ontologies that are mainly domain-specific and focused on adding support to relational statements rather than non-relational ones (Turki et al., 2019). Due to their triple-based structure, both knowledge graphs and ontologies can be queried using SPARQL - a query language ratified in 2008 by the World Wide Web Consortium (Angles and Gutierrez, 2008). However, property constraints allowing intrinsic data validation are differently represented in knowledge graphs and ontologies. Knowledge graphs are validated using Shape Expressions (ShEx) and the Shapes Constraint Language (SHACL) (Turki et al., 2022). As for ontologies, they are validated using the Semantic Web Rule Language (SWRL) rules (Liu et al., 2010).

Due to their extensible format, knowledge graphs are currently used in many fields including natural language processing (Schneider et al., 2022), explainable machine learning (Tiddi and Schlobach, 2022), and scholarly communication (Verma et al., 2022). With the rise of the Internet, open knowledge graphs have been released online under permissive licenses (e.g., *Creative Commons*) to provide access to large-scale structured and multi-disciplinary knowledge for people willing to develop knowledge-based systems using the semantic web principles (Färber et al., 2017). *DBpedia* (Created in 2007) is a knowledge base that is generated through mapping Wikipedia infoboxes, links, and categories as RDF triples using bot editors (Lehmann et al., 2015). *YAGO* (Created in 2008) is an ontological database that automatically aligns Wikipedia entries to WordNet synsets and GeoNames entities to generate multilingual items including multiple statements combining all the information included in Wikipedia infoboxes and categories (real-world data), WordNet (lexical data), and GeoNames (geographical data) (Rebele et al., 2016). *BabelNet* (Created in 2012) is a lexical database that automatically aligns language editions of the Wikipedia pages to WordNet synsets to generate multilingual items that exactly maintain the same relations as their WordNet equivalents (Navigli et al., 2021). *Wikidata* (Created in 2012) is a multilingual and multidisciplinary knowledge graph that is collaboratively edited by a community of volunteers and a set of bots and scripts allowing the crowdsourcing of structured knowledge from external semantic resources across the Linked Open Data Cloud (Vrandečić and Krötzsch, 2014). All these knowledge graphs are growing over the years and have proven their sustainability to remain operational across years, their scalability to grow in size, and their extensibility to cover new topics. They have also been upgraded to become Findable, Accessible, Interoperable, and Reusable (FAIR) using a wide range of programmatic tools (e.g., *APIs* and *data dumps*) and user interfaces (Turki et al., 2019).

## 3. Proposed Approach

As of March 26, 2023, we retrieved all the bibliographic metadata of scholarly publications related to open knowledge graphs between 2013 and 2022 from *Scopus*.[1]

After downloading the results, we manually reviewed them to isolate papers explicitly dedicated to open knowledge graphs. This blend of automated search and manual refinement ensured a focused and relevant dataset for our analysis. Papers that are just using open knowledge graphs to assess generic methods are consequently not included in this bibliometric study. Then, we analyzed the publication years, research venues, citation counts, and country affiliations for the considered publications and we identified the most cited research publications and the most commonly used keywords. Finally, to study the evolution of the topic coverage of scholarly research on open knowledge graphs, we construct the keyword co-occurrence networks for five different periods: *2013-2014*, *2015-2016*, *2017-2018*, *2019-2020*, and *2021-2022*.

---

[1]Scopus query: `TITLE-ABS-KEY ( "knowledge graph" OR "knowledge graphs" OR "knowledge base" OR "knowledge database" OR "semantic base" OR "semantic database" OR "knowledge basis" OR "graph database" OR "graph databases" ) AND TITLE-ABS-KEY ( "Open" OR "FAIR" OR "free" ) AND PUBYEAR > 2012`.

For this purpose, we apply VoSViewer, an open-source software for bibliographic mapping, to the *Author Keywords* and *Index Keywords* of the research publications as retrieved from Scopus (van Eck and Waltman, 2009). The visualization process assigns weight to the nodes according to their total link strength, normalizes the network clustering based on association strengths (Waltman et al., 2010), and discards keywords only included in less than eight research publications (van Eck and Waltman, 2009). As well, only the 1000 co-occurrence relations having the best association strengths in networks are visualized. The clustering of the nodes is based on the clustering technique of VOSviewer, a modularity-based clustering algorithm (Waltman et al., 2010). The resolution parameter in VOSviewer's clustering technique was set to the default value of 1.00. This parameter setting aligns with the guidelines provided in van Eck and Waltman (2014).

## 4. Time and Space Analysis

As revealed by Scopus, we identified 4445 scholarly articles utilizing open knowledge graphs published between 2013 and 2022. The yearly scholarly production on this topic has linearly evolved over the decade from 226 in 2013 to 751 in 2022 as shown in Figure 1. The research outputs for every year received an average of more than 4 citations per paper except the ones for 2020-2022 which have naturally fewer citations as they had less time to accumulate them.

The analysis of the citation counts for every single publication (Figure 2) shows that 31% of papers are never cited and two-thirds have 5 or fewer citations. On the other hand, about 12% of papers have accumulated 20 or more citations, suggesting that they have been noticed and used by the community. Finally, about 2% of papers became very popular, with over 100 citations. When seeing the list of the most cited publications (Table 1), we find out that they are descriptive papers for common open knowledge graphs, particularly multidisciplinary ones (e.g., *DBpedia* and *Wikidata*), biological ones (e.g., *Reactome pathway knowl-*

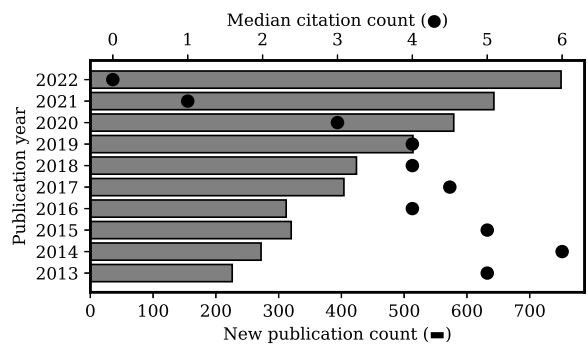

Figure 1: Citation and publication counts as dependent on the publication year.

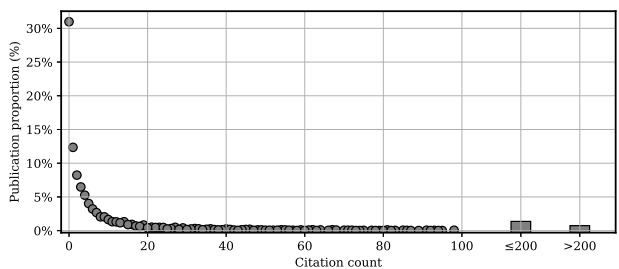

Figure 2: Distribution of accumulated citations of each paper.

*edgebase*) and medical ones (e.g., *ChestX-ray8*). Generic methods specifically developed for open knowledge graphs are consequently not very impactful and constitute a minority in the 2% most popular publications.

| Paper | Cited |
|---|---|
| Wang et al. (2016) | 1854 |
| Lehmann et al. (2015) | 1715 |
| Vrandečić and Krötzsch (2014) | 1617 |
| Wang et al. (2017) | 1541 |
| Croft et al. (2014) | 1176 |
| Pawson et al. (2014) | 777 |

Table 1: Most cited publications related to open knowledge graphs.

When analyzing the country distribution of the research publications (Figure 3), it appears that the publishing landscape is dominated by developed countries from North America and Europe (e.g., *United States of America*, *Germany*, *United Kingdom*, and *Italy*), Japan, South Korea, and BRICS (i.e., *Brazil*, *Russia*, *India*, and *China*) nations. When the research productivity of every country is normalized to its population size, it seems that developed countries from North America and Europe keep their prestigious place with a rate of five publications or more per one million inhabitants while Japan, South Korea, and BRICS countries are lagging behind them with a rate of 2.5 publications or less per one million inhabitants. Particular outliers in this perspective are the European countries of *Austria*, *Switzerland*, and the *Netherlands*.

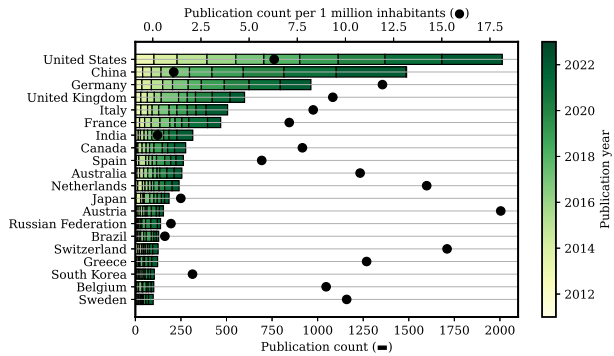

Figure 3: Citation and publication count as dependent on the country of affiliation. Demography data is retrieved from the *Countryinfo* Python Library.

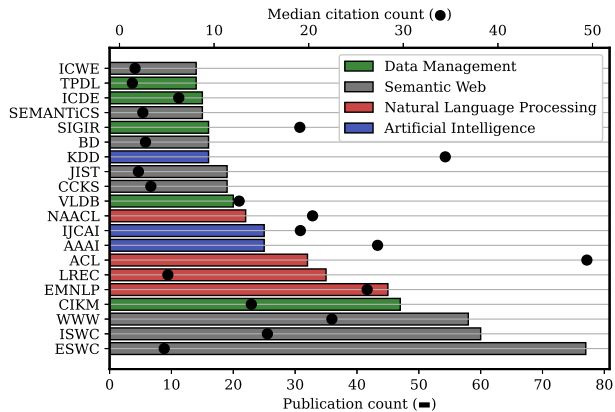

Figure 4: Citation and publication count as dependent on the conference names.

When identifying the conferences where publications related to open knowledge graphs are presented (Figure 4), we found a plethora of publication venues where such works are regularly submitted. These conferences are all of technical nature and not linked to other domains like Biomedicine, though some are oriented more towards text processing (e.g., *ACL*, *LREC*, and *EMNLP*), artificial intelligence (e.g., *AAAI* and *IJCAI*), data management (e.g., *CIKM* and *VLDB*), and some towards semantic web in general (e.g., *ESWC*, *WWW*, and *ISWC*). That being said, the top three conferences where open knowledge graph research is published are semantic web ones: *ESWC* (First), *ISWC* (Second), and *WWW* (Third). Despite the relative domination of semantic web conferences on the outputs of open knowledge graph research, the conferences where scholarly works on the topic are most impactful are Natural Language Processing (e.g., *ACL* and *EMNLP*) and Artificial Intelligence (e.g., *KDD*, *AAAI*) ones.

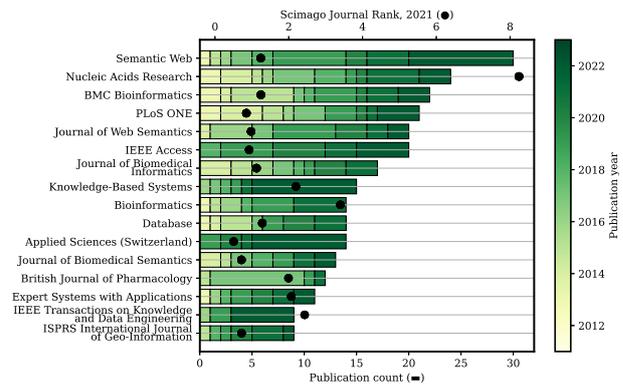

Figure 5: Scimago Journal Rank (SCImago, 2023) and publication counts for the most published scholarly journals.

When identifying the main scholarly journals publishing open knowledge graph research (Figure 5), we find four main categories of source titles. The first one is the set of application-oriented journals, featuring the applications of open knowledge graphs in fields beyond computer science like Biomedicine and mostly having a high citation impact (SJR ≥ 2). Examples of these journals are *Nucleic Acids Research*, *British Journal of Pharmacology*, *ISPRS International Journal of Geo-Information*, *BMC Bioinformatics*, *Bioinformatics*, *Database*, *Journal of Biomedical Informatics*, and *Journal of Biomedical Semantics*. The second group includes specialized scholarly journals about the Semantic Web, particularly *Semantic Web Journal* and *Journal of Web Semantics*. These journals do not have a very high citation impact (SJR ≤ 2). However, they publish a significant part of the research outputs of this research field. The third group is constituted of open-access mega-journals, mainly *PLoS One*, *IEEE Access*, and *Applied Sciences*. These journals are not very impactful (SJR ≤ 2). Finally, the fourth set concerns the sources publishing knowledge-based systems such as *Knowledge-Based Systems*, *Expert Systems with Applications*, and *IEEE Transactions on Knowledge and Data Engineering*. These journals have a high citation impact (SJR ≥ 2).

## 5. Keyword Analysis

When analyzing the main keywords included in open knowledge graph research, we find that these keywords include basic terminology of the semantic web and knowledge engineering fields, several renowned tasks for text processing, application fields for open knowledge graphs, and the names of hardware and software resources as shown in Figure 6. The basic terminology of the semantic web field like *semantic web*, *ontology*, *knowledge base*, *linked data*, and *graph database* has been featured in open knowledge graph research for all the period between 2013 and 2022. As for the other types of keywords, they can be classified based on their years of onset. The first period (2013-2016) is characterized by the use of open resources, including *Linked Open Data Cloud*, *Wikipedia*, and *open data* to construct open knowledge graphs. *DBpedia*[2] and *Wikidata*[3] are two examples of open knowledge graphs that have been initialized from relation extraction from Wikipedia and data integration across the Linked Open Data. At that time, the mainly used techniques range from classical machine learning algorithms to basic natural language processing (e.g., *named entity recognition*) and information retrieval (e.g., *text mining* and *relation extraction*) methods and data integration based on the semantic alignment between multiple resources. The main purpose of open knowledge graphs in this period is knowledge representation, particularly in the context of Big Data, and then data mining to find specific information inside the generated open resource using queries for question answering and customized recommendation purposes. In this context, *RDF* (2014), *SPARQL* (2015), and *Neo4J* (2016) have appeared as tools for enabling open knowledge graph hosting and querying. The second period (2017-2019) builds upon what has been achieved during the first period by defining new applications of open knowledge graphs, particularly digital humanities and cultural heritage, and using the recent advances of artificial intelligence, particularly deep learning to enhance the functionality of open knowledge graphs. Thanks to deep learning, new techniques and tasks have appeared like *link prediction* (2017), *knowledge graph completion* (2017), and *knowledge graph embedding* (2018). The third period (2020-2022) is mainly characterized by the beginning of the COVID-19 pandemic, a respiratory disease outbreak that began in 2020 and caused thousands of deaths, urging the move towards open science and the mass development and the increase of the scalability of open knowledge graphs so that they can become structured databases for fighting the disease outbreak and reduce its drawbacks on the worldwide human population.

Between 2013 and 2014, it is clear that scholarly research on open knowledge graphs was mainly application-oriented. As shown in Figure 7, there are three main keyword clusters. The one in blue reveals the existence of scholarly works on the development of knowledge-based systems driven by open-source software for natural language processing, information retrieval, knowledge engineering, computational linguistics, decision support, and semantic relation extraction where open knowledge graphs are both used as sources and targets of such intelligent systems. Furthermore, it highlights efforts for developing semantic web standards and languages for knowledge representation in open knowledge graphs. The red cluster shows significant interactions between the open knowledge graph applications in Chemistry, Medicine, Genetics, and Bioinformatics to build knowledge bases for OMICS research and medical education. It also finds a trend in utilizing open knowledge graphs for processing bibliographic databases for information retrieval and analyzing research production behaviours. As for the green cluster, it identifies the integration of open resources, particularly Ontologies, Wikipedia, and Linked Open Data Cloud, for constructing and enriching open knowledge graphs. It also indicates the development of interesting applications of knowledge-based systems for social network analysis and data mining in the medical context. Later, such systems can be useful to create customized open knowledge graphs reflecting the trends and evolution of the analyzed inputs. These clusters have significant co-occurrence associations as revealed by Figure 7, proving an important interdependence of the three topic clusters in achieving the final result: The creation and sustainability of open knowledge graphs.

---

[2]Research already started in 2013.

[3]Wikidata was created in 2012. Research about it only began in 2016.

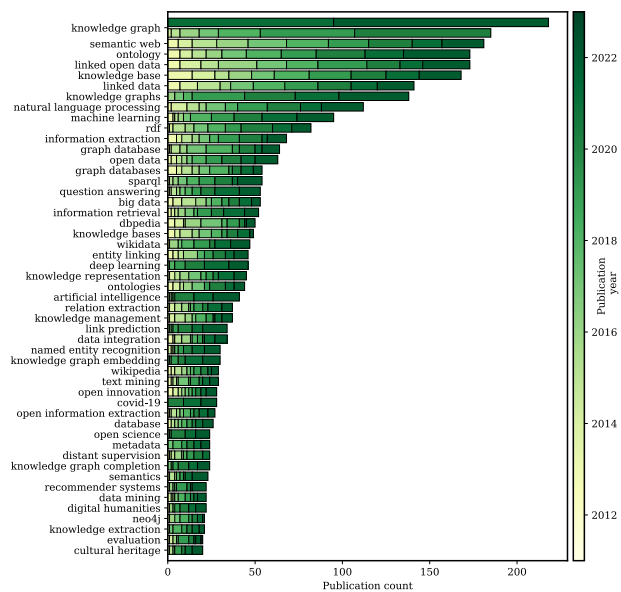

Figure 6: Most common keywords in scholarly publications related to open knowledge graphs. Note the split bars for *knowledge graph*.

When reproducing the network for 2015-2016 (Figure 8) and 2017-2018 (Figure 9), we find a similar behaviour of the topical coverage of scholarly research on open knowledge graphs. Again, we report a more compact integration of the open knowledge graphs applications for clinical medicine, molecular biology, and Bioinformatics to drive

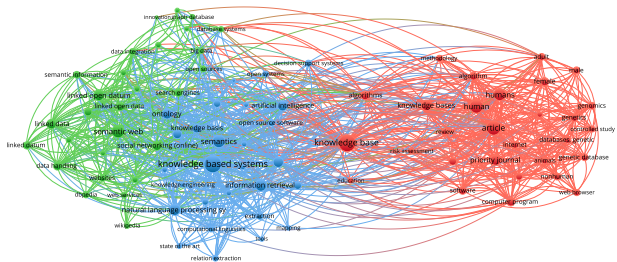

Figure 7: Keyword co-occurrence network for the research works published in 2013-2014.

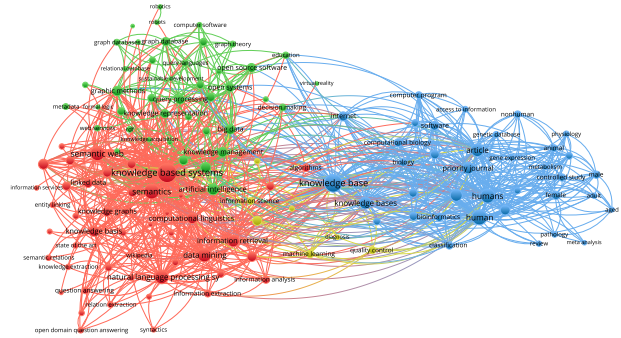

Figure 8: Keyword co-occurrence network for the research works published in 2015-2016.

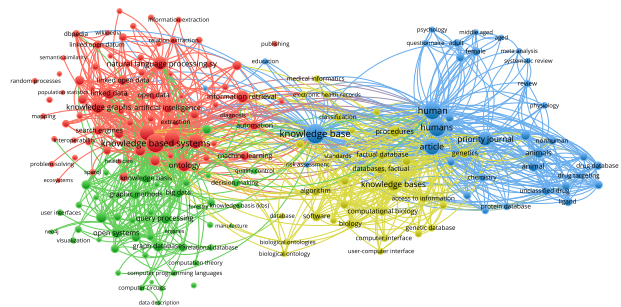

Figure 9: Keyword co-occurrence network for the research works published in 2017-2018.

OMICS research and medical education (Blue in Figure 8, Blue-Yellow in Figure 9). As in 2013-2014, this cluster also includes the extraction of structured data from scholarly publications to enrich and validate open knowledge graphs. The only new update has been the development of algorithms for the construction of structured data from electronic health records to automate disease and medical procedure management since 2017-2018 (Yellow in Figure 9). Furthermore, we identify the stability of the efforts towards the development of intelligent systems based on open knowledge graphs for various purposes including natural language processing, data mining, and information retrieval with a better emphasis on new topics like question answering and demography.[4] We also identify the introduction of a new trend of using machine learning algorithms at the same period for training knowledge-based systems based on open knowledge graphs and for enriching and validating open knowledge graphs (Yellow in Figure 8, Red in Figure 9). It is clear as well that the trend towards data integration of free resources (e.g., *Wikipedia* and *Linked Open Data*) and semantic web standards for constructing and enriching open knowledge graphs, allowing to prevent the cold start problem in initiating such databases, is still ongoing.[4] Moreover, we report a new and significantly growing tendency for developing a hardware and software infrastructure to host and process open knowledge graphs (Green in Figures 8 and 9). This includes the development of query languages (e.g., *SPARQL*) and validation ones (e.g., *ShEx*), the optimization of query endpoints, and the construction of big data infrastructures for large-scale open knowledge graphs. This effort is mainly driven by the use of semantic web standards (e.g., *RDF*) and open-source software (e.g., *Neo4J*) and is supported by a theoretical work on upgrading the functionality of open knowledge graphs based on computation theory and graph theory. This work has partially been done to support the emergence of open knowledge graphs for new applications that require less runtime and more advanced data complexity such as

Robotics, Education, Virtual Reality, Web Services, and Industry.

The generation of the keyword co-occurrence network for 2019-2020 (Figure 10) and 2021-2022 (Figure 11) confirms the trends of open knowledge graph research towards the development of knowledge-based systems for various applications (Red in Figures 10 and 11), the integration of free resources for supporting open knowledge graphs,[5] the applications of open knowledge graphs in Clinical Medicine, Medical Education, Bioinformatics, and information retrieval from scholarly publications (Green in Figures 10 and 11), and the development of algorithms for sustainable query processing and optimization.[5] The only difference in the efforts towards data integration is the emphasis on the Findability, Accessibility, Interoperability, and Reusability (FAIR) principles of open knowledge graphs, implying the resolution of multiple legal and technical barriers (Blue in Figures 10 and 11). As for the medical applications of open knowledge graphs, what differs is the interest of the research community in the COVID-19 pandemic (e.g., *COVID-19*, *coronavirus disease 2019*, and *SARS-CoV-2*). Beyond this, major differences between the trends in open knowledge graph research before and after 2019 are the development of open knowledge graphs for cultural heritage and digital humanities as well as for Internet of Things-driven Industry (Blue in Figure 10), the use of open knowledge graphs as an input and output for deep learning techniques (e.g., *long short-term memory*) and language models (e.g., *embeddings*) (Red-Yellow in Figure 10, Red in Figure 11) allowing the generation of novel machine learning algorithms for knowledge graph processing (e.g., *graph embeddings* and *graph neural network*), and the development of open knowledge graphs including lexicons and morphology data for natural languages and data about social interactions across heterogeneous social networks (Purple in Figure 11). The latter serves for language processing and social network analysis and requires robust cybersecurity algorithms for ensuring network security and preventing adversarial attacks (Purple in Figure 11). The purple cluster emphasizes the interconnection between trustworthy artificial intelligence and machine learning in the context of knowledge graph creation and development (Tiddi and Schlobach, 2022; Schramm et al., 2023). This involves using knowledge graphs to elucidate machine learning outputs (Tiddi and Schlobach, 2022) and demonstrating the application of trustworthy machine learning in the evolution of knowledge graphs (Schramm et al., 2023). Trustworthy machine learning is grounded in the development of algorithms designed to ensure that the output of machine learning models is ethical, robust, responsible, explainable, secure, and fair, fostering a comprehensive approach to AI development (Liu et al., 2022). The only facets covered by the cluster are robustness and security, proving that more efforts should be provided by the scientific community to ensure that research about machine learning for knowledge graph development contributes to society and human well-being, avoids harm, and upholds professional standards of conduct and ethical practice.

---

[4] Red in Figures 8 and 9.

[5] Blue in Figure 10, Yellow in Figure 11.

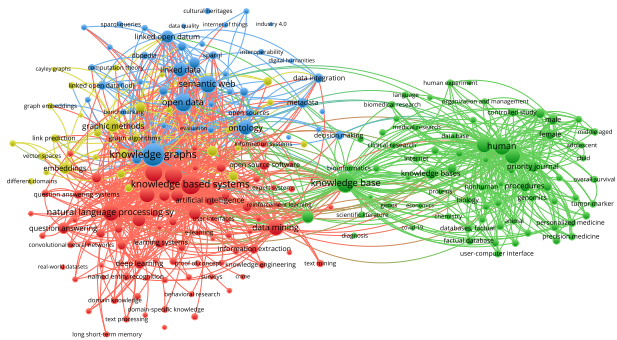

Figure 10: Keyword co-occurrence network for the research works published in 2019-2020.

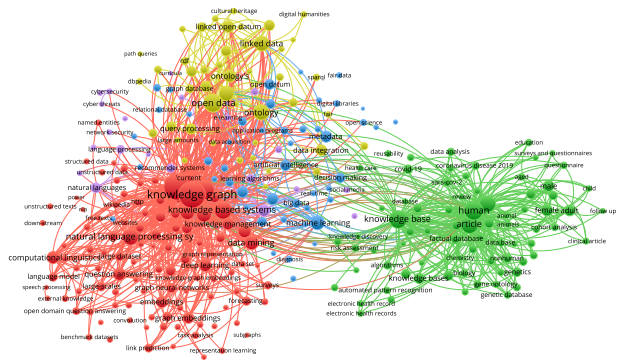

Figure 11: Keyword co-occurrence network for the research works published in 2021-2022.

## 6. Discussion

The increasing productivity of research outputs about open knowledge graphs (Figure 1) reflects the growth of interest of the scientific community in such open resources. With the development of open science, particularly following the COVID-19 pandemic (Homolak et al., 2020), open knowledge graphs are more involved as resources for reproducible and trustworthy knowledge-based artificial intelligence (Edelstein et al., 2020). As well, the rise of open knowledge graph research is significantly linked to the evolution of scholarly research on the semantic web (Gandon, 2018) and knowledge graphs (Hogan et al., 2021) over the years. The stable rate of around 4 citations per publication for open knowledge graph research is acceptable when compared to ones of other research fields like economics (Li and Ho, 2008) and chemistry (Yi et al., 2008) and confirms the interest of the scientific community in this research field. The distribution of citations between publications where a few papers receive a significant number of citations (Figure 2) goes in line with the shifted Lotka function that characterizes how citations are allocated to a set of scholarly publications (Egghe and Rousseau, 2011). The higher citation impact of the description papers of open knowledge graphs (Table 1) is quite surprising as the most-cited publications in computer science are mainly literature reviews or widely used generic algorithms (Ha, 2022). When seeing the source titles for all the considered research publications, we find that the target journals for open knowledge graph research are quite the same as for knowledge graphs research in general where major scholarly journals in knowledge engineering, semantic web, and applied informatics as well as open-access mega-journals are featured as the most published source titles (Chen et al., 2021). The only difference is the existence of *Database*, a specific journal for open data curation, among the top publication venues for open knowledge graph research. As for scholarly conferences, they are mainly CORE A or A* highly-referred ones except for several CORE B (*BigData* [BD], *TPDL*, and *ICWE*), CORE C (*LREC*), and unclassified (*CCKS*, *JIST*, and *SEMANTICS*) scholarly conferences (Padgham et al., 2021). These CORE A* and A venues are considered the most important scholarly conferences for knowledge engineering (Vahdati et al., 2020). Publications in these highly-referred conferences except *ESWC* have a citation rate that is largely above the average ($\geq$ 4 citations per publication) while other conferences have a limited citation impact. This confirms the tendency of CORE A* and A conferences to include more impactful research publications than other sources due to their emphasis on timely innovation-driven research that provides game-changing breakthroughs (Vrettas and Sanderson, 2015).

The most published countries are mainly nations with high traditions in knowledge graph research (Chen et al., 2021), led by the *United States of America*, *China*, and *Germany*. These countries can be classified into two categories (Figure 3): European and North American developed countries, and other countries, including fast-growing mostly over-populated countries including BRICS (*Brazil*, *Russia*, *India*, and *China*), *South Korea*, and *Japan*. European and North American countries have a high rate of publications per capita ($\geq$ 2.5 per one million inhabitants) by contrast to other most productive countries. This fact is mainly explained by the better funding provided to researchers in Europe and North America (Shashnov and Kotsemir, 2018). The topics of open knowledge graph research, emphasizing the development of knowledge-based systems based on these resources (Gandon, 2018), the applications of open knowledge graphs in natural language processing, data mining, and biomedicine (Schneider et al., 2022; Gandon, 2018; Ristoski and Paulheim, 2016), and the integration of open resources for generating open knowledge graphs (Mountantonakis and Tzitzikas, 2019) (Figures 7 to 11), are common in the overall knowledge graph research. The occurrence of the *COVID-19 pandemic* in 2020 encouraged the development of these three topics in practice (Chatterjee et al., 2021). The emphasis of *Wikidata* and *DBpedia* as large-scale open knowledge resources (Table 1) is mainly due to the existence of a research and development community that collaboratively contributes to these two knowledge graphs (Mora-Cantallops et al., 2019; Färber et al., 2017). The shift towards establishing new applications of open knowledge graphs in fields like *industry* (2017-2018), *cultural heritage* (2019-2020), and *applied linguistics* (2021-2022) and the use of new techniques, particularly *query processing* (2015-2016), *logical reasoning* (2015-2016), *knowledge graph embeddings* (2019-2020), *graph neural networks* (2021-2022), and *network security* (2021-2022) simultaneously occurred with the general trends of knowledge graph representation and reasoning (Cambria et al., 2021), learning (Chen et al., 2021), and processing (Gandon, 2018; Hogan et al., 2021). However, it seems that these aspects appeared in open knowledge graph development before scholarly research. Query (e.g., *SPARQL*) and validation (e.g., *ShEx* and *SHACL*) languages have been created between 2008 and 2017 and used in practice since that period (Gandon, 2018; Xue and Zou, 2022; Paulheim, 2016). The development of lexical knowledge graphs began before 2020 with large knowledge resources like *BabelNet* (Navigli et al., 2021) (Created in 2011) and *Wikidata lexicographical data* (Nielsen, 2019) (Created in 2018). The use of relational machine learning for constructing and validating knowledge graphs is also an old topic that has been transformed thanks to the advances in deep learning and pre-trained models (Xia et al., 2021). Such behavior is not common in the other computer science subfields such as machine learning where scholarly research is ahead of the industry (Kumar et al., 2020).

## 7. Summary

In this research paper, we focus on studying the quantitative evolution of scholarly research on open knowledge graphs between 2013 and 2022. We have observed that works in the field is becoming increasingly important and new concepts are emerging. Since 2019, there has been more interest in developing the interaction between open knowledge graphs and advanced machine-learning techniques for better coverage and quality of these open resources. However, we have observed that the development of breakthroughs in the field begins with development and industry applications instead of research projects. This should encourage the community to rethink the triple helix relation (Leydesdorff and Deakin, 2011) between government, research, civil society, and industry in open knowledge graph research.

The enhancement of open knowledge graph research can be enabled in the next few years by encouraging cross-disciplinary research to integrate open knowledge graphs into various scientific disciplines. Future directions for this work can be the extension of this bibliometric study to include a broader range of data sources, such as other academic databases, preprint servers, and grey literature, to provide a more comprehensive analysis of the field. We also envision incorporating qualitative analysis, such as interviews or surveys with researchers in the field, to provide a more nuanced understanding of the research themes and directions.

## 8. Data Availability

For reproducibility purposes, source code and data are made available at `https://github.com/Data-Engineering-and-Semantics/OpenKGBiblio/`.

## 9. Acknowledgments

This work is a part of the *Adapting Wikidata to support clinical practice using Data Science, Semantic Web and Machine Learning* project funded by the Wikimedia Research Fund, an initiative of the Wikimedia Foundation. We thank Vilém Zouhar for help with the analysis and visualization aspects of this work.

## 10. Bibliographical References